\documentclass{collective-intelligence}

\usepackage{graphicx}
\usepackage{textcomp}
\usepackage{color}

\begin{document}
\bibliographystyle{agsm}

\title{Analytic Methods for Optimizing Realtime Crowdsourcing}
%
%
%
%
%

\numberofauthors{2} 
%
\author{
%
%
\alignauthor
Michael S. Bernstein, David R. Karger, Robert C. Miller\\
       \affaddr{MIT CSAIL}\\
       \affaddr{Cambridge, MA 02139}\\
       \email{\{msbernst, karger, rcm\}@csail.mit.edu}
\alignauthor
Joel Brandt\\
       \affaddr{Adobe Systems, Advanced Technology Labs}\\
       \affaddr{San Francisco, CA 94103}\\
       \email{joel.brandt@adobe.com}
}

\maketitle
\begin{abstract}
Realtime crowdsourcing research has demonstrated that it is possible to recruit paid crowds within seconds by managing a small, fast-reacting worker pool. Realtime crowds enable crowd-powered systems that respond at interactive speeds: for example, cameras, robots and instant opinion polls. So far, these techniques have mainly been proof-of-concept prototypes: research has not yet attempted to understand how they might work at large scale or optimize their cost/performance trade-offs. In this paper, we use queueing theory to analyze the \emph{retainer model} for realtime crowdsourcing, in particular its expected wait time and cost to requesters. We provide an algorithm that allows requesters to minimize their cost subject to performance requirements. We then propose and analyze three techniques to improve performance: push notifications, shared retainer pools, and \emph{precruitment}, which involves recalling retainer workers before a task actually arrives. An experimental validation finds that precruited workers begin a task 500 milliseconds after it is posted, delivering results below the one-second cognitive threshold for an end-user to stay in flow.
\end{abstract}

\section{Introduction}
Crowdsourcing is no longer constrained to slow, offline tasks. Just as traditional programming evolved from offline batch processes to realtime results and interaction, crowdsourcing is now transitioning from wait times of hours \cite{ipeirotis2010analyzing} to seconds \cite{twoseconds}. Techniques that place workers on active retainer can now recruit crowds in two seconds \cite{twoseconds}, complete traditional crowdsourced votes in five seconds, and maintain continuous control of remote interfaces \cite{Lasecki2011}. These realtime crowds open the door to deployable applications that react intelligently, including smart cameras, robot navigators, spreadsheets, and on-demand graphic design.

However, existing realtime techniques have largely been prototypes aimed at demonstrating feasibility---they did not attempt to understand how these approaches would work at large scale or to optimize cost/performance trade-offs. We focus on the \emph{retainer model}, which pays workers a small extra wage to be on call while they pursue other tasks, then respond quickly when a realtime request arrives \cite{twoseconds}. Currently, the retainer model is not optimized for cost or performance, nor do requesters have any analytic framework to understand the relationship between retainer pool size, cost, and response time.

This paper analyzes the retainer model using queueing theory~\cite{gross1998fundamentals} to understand its performance at scale, in particular the trade-off between expected wait time and cost. We introduce a simple algorithm for choosing the optimal size of the retainer pool to minimize total cost to the requester subject to the requester's performance requirements: maximum expected wait time or maximum probability of missing a request. We then propose several improvements to the retainer model that reduce expected wait time. First, \emph{retainer subscriptions} allow workers to sign up for push notifications for recruitment, which reduces the length of time it takes to recruit new workers onto retainer. Second, \emph{combining retainer pools} across requesters leads to both cost and wait time improvements. Large retainer pools can then be made more effective by using \emph{task routing} to connect appropriate workers to the tasks that need them. Third, a \emph{precruitment} strategy recalls workers from retainer a few moments before a task is expected to arrive, dramatically lowering response time. We perform an early empirical evaluation demonstrating that precruitment results in median response times of just 500 milliseconds.

Our analysis carries several benefits. First, realtime tasks can now directly minimize their cost for a given performance requirement. Second, the retainer subscriptions allows workers to register for the tasks they like best and have them delivered, rather than constantly seeking out new work. Third, we demonstrate empirically that these techniques can overcome previous limits of ``crowds in two seconds'' to deliver the feedback to the user within 500 milliseconds---finally under the one-second cognitive threshold for an end-user to remain in flow \cite{nielsen1993usability}.

We begin by surveying related work on realtime crowdsourcing and wait times in crowdsourcing systems. We then describe the retainer model and use queueing theory to analyze and optimize wait time and cost. We introduce our improvements to the model---retainer subscriptions, global retainer pools, and precruitment---and integrate them into our analysis. Finally, we discuss limitations of our approach and point to future work realizing the vision of realtime crowds.

\section{Related Work}
Crowdsourcing researchers have a strong interest in fast task completion times. Paying more will lead to more work completed \cite{Mason2009}, but not at realtime speed. QuikTurKit introduced two techniques to improve response time: repeatedly posting tasks so as to stay visible in the recent task list, and keeping workers primed through old tasks until a new task is ready \cite{vizwiz}. The retainer model builds on QuikTurKit by paying workers a small fee and notifying them when work is ready, recruiting crowds in two seconds \cite{twoseconds}. Workers can also maintain continuous realtime control of an interface by electing temporary leaders \cite{Lasecki2011}. We contribute a more thorough analysis of the techniques in these systems and an algorithmic approach to helping these systems achieve target wait times at minimum cost.

Accurate models of crowdsourcing platforms help us understand the underlying processes and predict behavior when parameters change. Queueing theory has been used to estimate throughput and wages on Mechanical Turk \cite{ipeirotis2010analyzing}. Survival analysis is another popular model for predicting task completion time, especially in non-realtime scenarios \cite{Faridani2011}. We model crowdsourcing task arrival processes as Poisson; empirical data suggests that the Poisson approximation is accurate when parametrized by time of day \cite{Faridani2011}.

\section{Retainer Model}
The retainer model is a recruitment approach for realtime crowdsourcing. This model was introduced for realtime interfaces like instant feedback votes and a crowdsourced camera shutter \cite{twoseconds}. It pays workers a small wage to be on call and return quickly when a task is ready. These workers accept the task in advance and are paid extra to keep their browser window open. While they wait for the task to arrive, workers are free to work on other tasks. There are many methods for recalling the worker; \citeasnoun{twoseconds} used a modal dialog and an audio alert. Evaluations demonstrated that workers messaged in this way begin work in two to three seconds.

\section{Queueing Theory Analysis}
In this section, we investigate a mathematical model of retainers. This model allows us to predict how long realtime tasks will need to wait. To begin, suppose each task type has its own set of retainer workers. When a task comes in, a worker leaves the retainer pool to work on the task and the retainer system recruits another worker to refill the pool. The goal is to maintain a large enough pool of retainer workers to handle incoming tasks. In other words, we want to minimize the probability that the retainer pool will be empty (no retainer workers left), subject to cost constraints. The risk is that a burst of task arrivals may exhaust the retainer pool before we can recruit replacement workers.

We will model this problem using queueing theory. In queueing theory, a set of \emph{servers} are available to handle \emph{jobs} as they arrive. If all servers are busy handling a job when a new job arrives, that job enters a queue of waiting tasks and is serviced as soon as it reaches the front of the queue. In our scenario, tasks are jobs, and retainer workers are servers.

In this paper, we will consider a class of algorithms that set an optimal retainer pool size. Suppose the retainer pool is $c$ workers.  As jobs come in and remove workers from the retainer pool, assume that the system always puts out enough requests for new workers to bring the pool back to $c$. That is, if there are $c_0$ workers in the pool, the system has issued $c-c_0$ outstanding requests. If, when a job arrives, the pool is empty, the system sets it aside for special processing: it directly recruits a worker, not for the pool, but for that job. In effect, a user with a diverted job is immediately alerted that the system is over capacity and the job will be handled out-of-band after a short delay. This final assumption may not accurately reflect how a running system would work, but it provides an upper bound on expected wait time and makes it easier to analyze the probability that a task will be serviced in realtime.

Suppose that tasks arrive as a Poisson process at rate $\lambda$, and retainer workers arrive after they are requested as a Poisson process at rate $\mu$.\footnote{These assumptions are perhaps overly ideal. Job arrivals on Mechanical Turk are heavy-tailed \cite{ipeirotis2010analyzing}. However, much of our analysis is independent of the arrival distribution, and systems can always substitute empirically observed distributions and solve numerically.} Then, the empty spots in the retainer pool, each of which will become filled when a worker arrives, can be thought of as busy machines occupied with a job whose completion time is a Poisson process with rate $\mu$. In our setup, we also divert jobs that arrive when all machines are busy.

In other words, this is an M/M/c/c queue where jobs arrive at rate $\lambda$ and have processing time $\mu$. A basic M/M/1 queue assumes Poisson arrival and completion processes, a single server, and a potentially infinite queue. An M/M/c/c queue has $c$ parallel machines instead of one, and rejects or redirects requests when there are no servers to immediately handle the incoming request \cite{gross1998fundamentals}. Imagine a telephone system, for example, that gives a busy signal if all $c$ lines are busy. The meaning of $\mu$ has changed slightly to indicate worker recruitment time instead of a job completion time, but the mathematical analysis is the same.

To optimize performance, we need to understand the probability that all workers are busy, since that is the case where a job has to wait (for expected time $1/\mu$). We also need to understand the cost of having a retainer pool of size $c$. Since the system pays workers proportional to how long they are on retainer without a job, the total cost is proportional to the average number of \emph{idle} machines---these are the ones representing workers waiting on retainer. Finally, we will eventually need to integrate worker abandonment into our model, since not all workers respond to the retainer alert.

\subsection{Probability of an Empty Pool}
The probability that a job must wait can be derived using Erlang's loss formula \cite{gross1998fundamentals}. We set $\rho$, the \emph{traffic intensity}, to be the percentage of system resources that are are required to service newly incoming tasks:  $\rho = \lambda/\mu$. In M/M/c/c queueing systems, as we will demonstrate, $\rho < c$ is necessary for the system to keep up with incoming requests.

The probability of an empty retainer pool (all $c$ ``servers'' busy) is Erlang's loss formula:
\begin{equation}
\label{eqn:erlang}
\pi(c)=\frac{\rho^c/c!}{\sum_{i=0}^{c} \rho^i/i!}
\end{equation}
A remarkable property of Erlang's loss formula is that this relationship requires no assumptions about the distributions of job arrival time or worker recruitment time, in particular whether they are Poisson.  It only depends on the means $\mu$ and $\lambda$.

\subsection{Expected Waiting Time}
For some applications, the probability of a task needing to wait is less important than the expected wait time for the task. The two quantities are directly related. The expected wait time is the probability of an empty retainer pool multiplied by the expected wait time when the pool is empty, or $\frac{1}{\mu}\pi(c)$:

\vspace{-.2in}
\begin{equation}
\label{eqn:expectedwait}
\frac{1}{\mu}\pi(c) = \frac{1}{\mu}\frac{\rho^c/c!}{\sum_{i=0}^c \rho^{i}/i!}
\end{equation}
This expression gives us a direct relationship between the size of the retainer pool, the arriving task and worker rates, and the expected wait time.

\begin{figure*}
\centering
\resizebox{\textwidth}{!}{\includegraphics{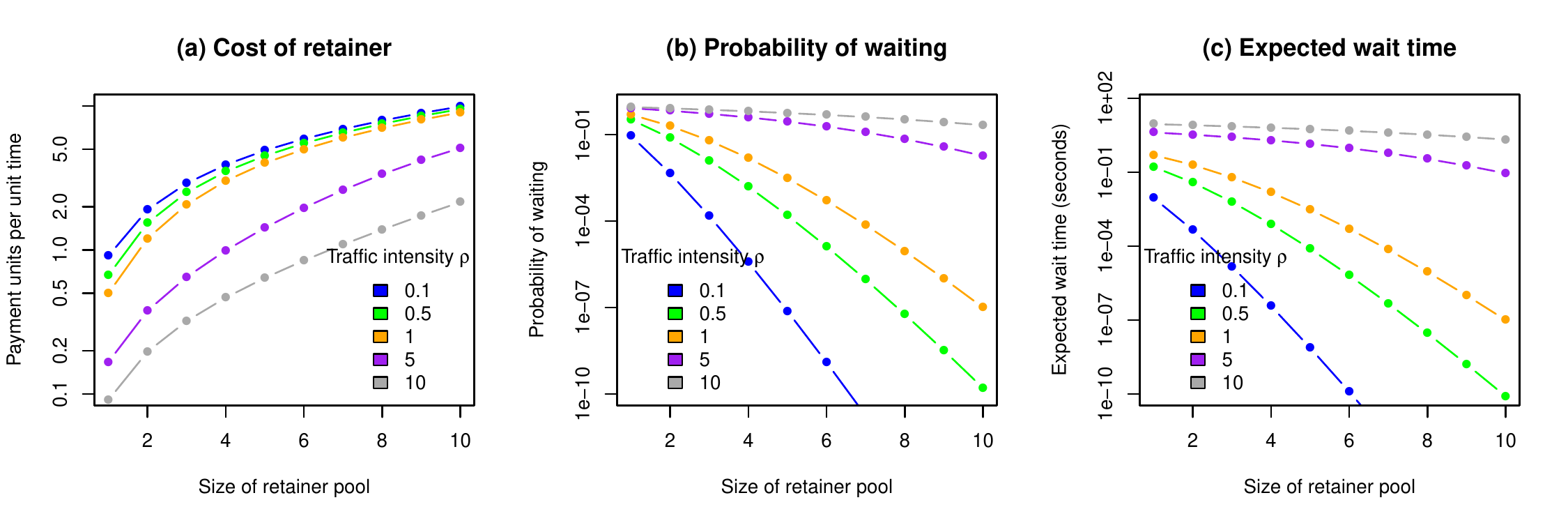}}
\vspace{-25pt}
\caption{Graphs that visualize the relationships between retainer pool size, traffic intensity, and (a) cost, (b) probability of a task waiting, and (c) expected wait time. In the graph of expected wait time, we set $\lambda=1$, so $\mu=\rho^{-1}$. When $\rho>c$, there are often not enough workers on retainer to service all tasks. As a result, wait time goes up, but cost goes down.}
\vspace{-10pt}
\label{fig:equations}
\end{figure*}

As a sanity check: when $\lambda \rightarrow 0$ (few arrivals) we have $\rho \rightarrow 0$ in which case $\pi(c) \rightarrow 0$.\footnote{Actually, $\pi(c) \rightarrow \rho^c/c!$} In other words, we are very unlikely to have an empty pool so the expected wait time also goes to zero. This relationship is visualized in Figure~\ref{fig:equations}(c).

\subsection{Expected Cost}
Once we understand expected waiting time, we can analyze the retainer model's cost characteristics. \possessivecite{twoseconds} experiments suggested that workers could be maintained on retainer for \$0.30 per hour at a rate of $\frac{1}{2}$\textcent~per minute, but this analysis is fairly simplistic. To understand cost more completely, we need to know the expected number of workers on retainer.

The probability of having $i$ busy servers in an M/M/c/c queue is a more general version of Erlang's loss formula:
\begin{align}
\pi(i)=\frac{\rho^i/i!}{\sum_{i=0}^{c} \rho^i/i!}
\end{align}
We can derive the closed form expression of the expected number of busy servers:
\begin{align}
E[i]  &= \frac{\sum_{i=0}^{c} i \rho^i/i!}{\sum_{i=0}^{c} \rho^i/i!}\notag\\
&= \rho \frac{\sum_{i=0}^{c-1} \rho^i/i!}{\sum_{i=0}^{c} \rho^i/i!}\notag\\
&= \rho (1-\pi(c))\label{eqn:busyservers}
\end{align}

In steady state, we need to pay all retainer workers who are \emph{not} busy. That is, we expect to have $c - \rho (1-\pi(c))$ workers waiting on retainer. If our retainer salary rate is $s$ (e.g., $s=\frac{1}{2}$\textcent), we would pay $s(c - \rho (1-\pi(c)))$ per unit time on average.

\subsection{Visualizing the Relationships}
While these equations give us precise relationships, they may not convey intuitions about the performance of the platform. Figure~\ref{fig:equations} plots these relationships for several possible values of $\rho$. These figures show a knee in the curve at $c \approx \rho$ for getting a good probability of response. A pool size $c>\rho$ means that an empty pool's \emph{overall} rate of recruitment of workers, $c\mu$, exceeds the arrival rate of tasks. In other words, we begin to catch up and rebuild a set of available workers.\footnote{When $\rho/c \rightarrow 0$, the number of free workers goes to $c-\rho(1-\rho^c/c!)$, or effectively $c$.}  On the other hand, if $c<\rho$, then even an empty queue will not recruit workers fast enough to cover all arriving tasks, so it will stay empty.\footnote{As $\rho \rightarrow \infty$, the number of free workers goes to $c-\rho c/(\rho+c)=c(1-\rho/(\rho+c))$ which goes to 0.}

\begin{figure}
\centering
\vspace{-10pt}
\resizebox{\columnwidth}{!}{\includegraphics{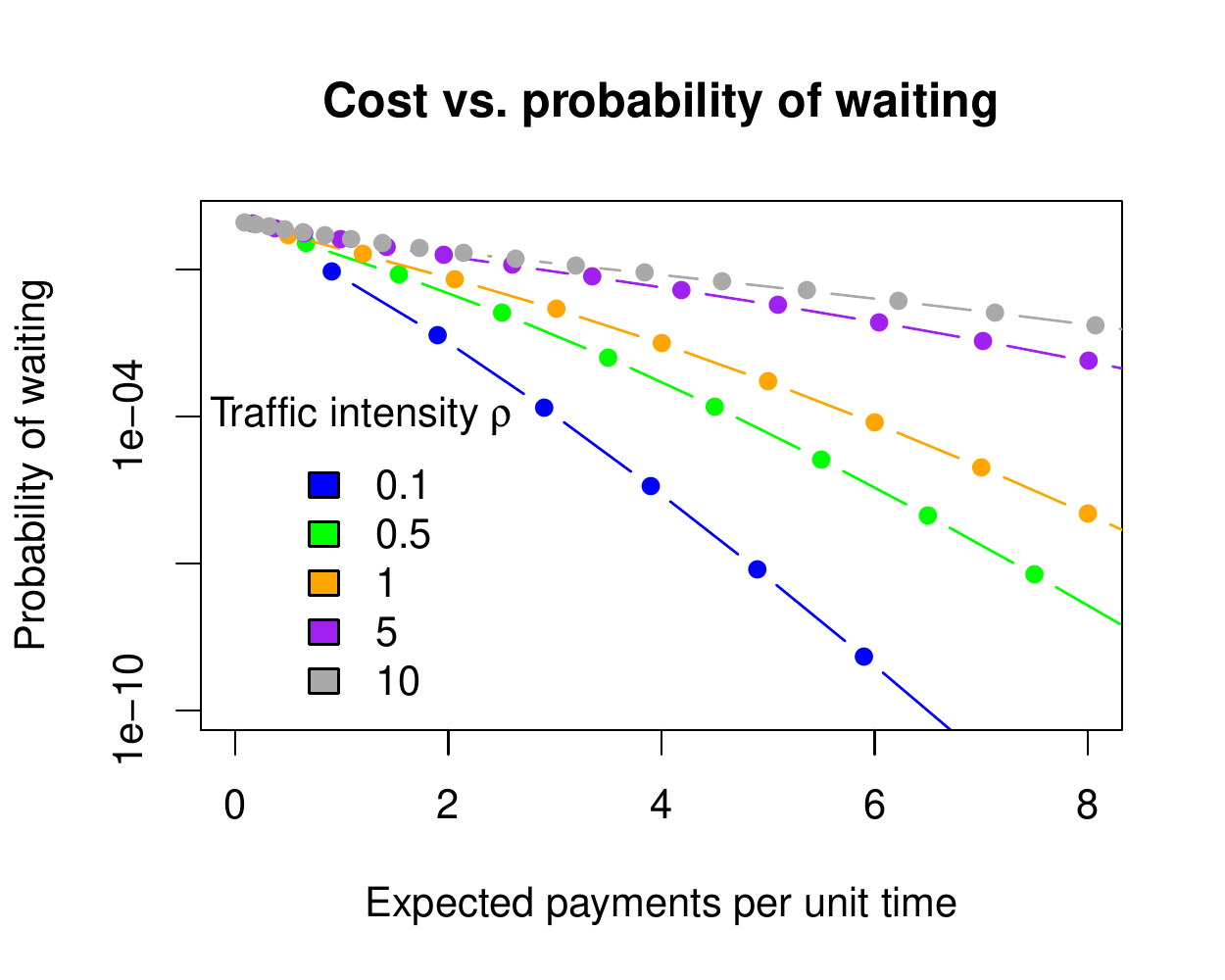}}
\vspace{-25pt}
\caption{By calculating cost and the probability of a task needing to wait for integer values of $c \in [1,15]$, we can visualize the relationship between the two values.}
\vspace{-20pt}
\label{fig:tradeoff}
\end{figure}

Figure~\ref{fig:tradeoff} visualizes the relationship between the requester's cost and the probability of waiting. We derive this parametric curve by choosing values of $c$, then finding the cost and probability of waiting given that value. Paying more (i.e., for a larger pool) always improves the probability that the system can immediately handle a request. However, for small values of $\rho$, e.g. $\rho \le 1$, paying 1--1.5\textcent~per minute brings the probability of waiting near zero. When tasks arrive quite quickly, 2.5\textcent~or more is necessary to achieve similar performance.

\section{Optimal Retainer Pool Size}
A queueing theory model allows us to determine the number of workers to keep on active retainer. The size of the retainer pool is typically the only value that requesters can manipulate, and it impacts both cost and expected wait time. Requesters want to minimize their costs by keeping the retainer pool as small as possible while also maintaining a low probability that the task cannot be served in realtime. In this section, we present techniques for choosing the size of the retainer pool.

Our goal is to find an optimal value of $c$, given 1) the arrival rates $\lambda$ and $\mu$, and 2) desired performance, in terms of the probability of a miss $\pi(c)$ or total cost. We assume that the requester knows $\lambda$ and $\mu$ either through empirical observation or estimation. We also assume that $\lambda$ and $\mu$ are constant, but it is enough just for them not to change too quickly.

One approach to finding $c$ is to specify the maximum allowable expected wait time for a task, or (equivalently) the maximum allowable probability that an incoming task will not be served in realtime. The intuition for this approach can be seen in Figure~\ref{fig:equations}(b): if $\rho=.5$, for example, and the requester wants a less than 5\% probability of any given task needing to wait, then $c=3$ is the smallest retainer pool that can make such a guarantee.

Algorithmically, if $p_{\text{max}}$ is the maximum desired probability of a task not being served in realtime, we want to minimize $c$ subject to $\pi(c) \le p_{\text{max}}$. To find the solution, we use a binary search over possible values of $c$.

A more interesting version of the problem is for the requester to attach a dollar value to each task that cannot be serviced in realtime. For example, some pizza delivery companies do not charge the customer for the pizza if they cannot deliver it within thirty minutes. A miss then costs the company the value of the pizza plus the deliveryman's wage spent delivering the late pizza. A requester might similarly offer the service for free if it is not completed in realtime, or they might decide that the bad experience of a non-realtime result is worth \$1 in lost potential revenue from that user.

\begin{figure}
\centering
\vspace{-10pt}
\resizebox{\columnwidth}{!}{\includegraphics{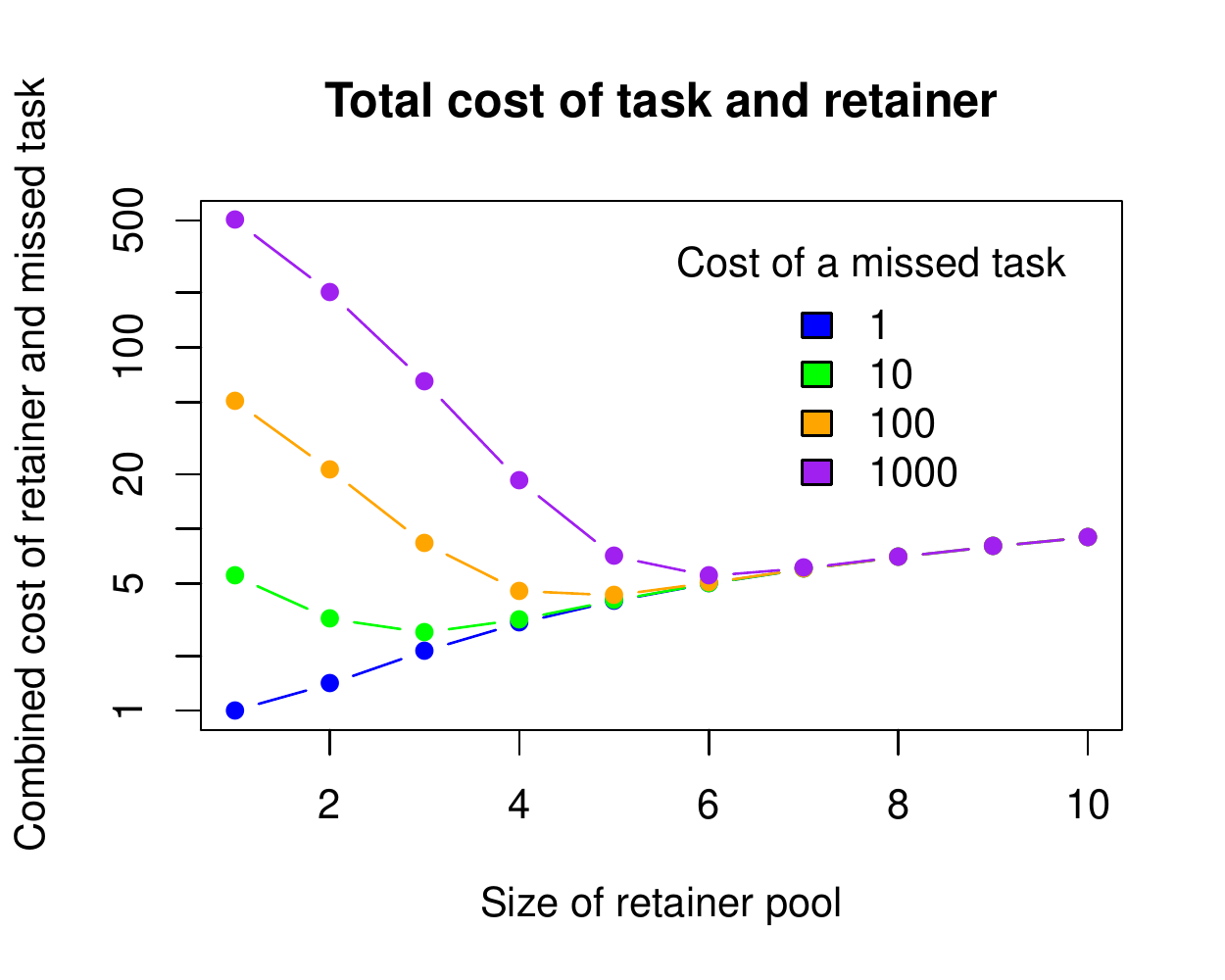}}
\vspace{-25pt}
\caption{By assigning a dollar value to missed tasks, we can visualize the relationship between retainer size and total cost. Assuming traffic intensity $\rho=1$ and retainer wage $s=1$, these curves demonstrate the trade-off between more missed tasks on the left part of the graph and higher retainer costs on the right.}
\label{fig:combinedcost}
\vspace{-15pt}
\end{figure}

It now becomes possible to directly minimize the requester's total cost. Let $C_{total}$ be the expected total cost to the requester and $C_{task}$ be the loss if a task is not completed in realtime. Then $C_{total}$ is the sum of the expected task cost---zero if addressed in realtime, or $C_{task}$ otherwise---and the wage for the retainer workers derived from Equation \ref{eqn:busyservers}:
\begin{equation}
\label{eqn:totalcost}
C_{total} = C_{task}\pi(c) + s(c - \rho (1-\pi(c)))
\end{equation}
We can minimize this total value. Figure~\ref{fig:combinedcost} shows this curve for several possible values of $C_{task}$ when $\rho=1$. The minimum value on the y axis for each curve is the optimal retainer size.

\section{Worker Abandonment}
Queueing theory models assume that a server will always begin a job once it is assigned. However, workers will sometimes leave the computer, close the window, or otherwise not respond to the retainer alert. Empirically, \citeasnoun{twoseconds} found that about 10-20\% of workers on active retainer never responded.

Our model can be adapted to capture worker abandonment. Let $a$ be the percentage of workers who abandon the task, that is, they do not return after the retainer alert. A straightforward edit is to add the constant $a$ to the probability that a task will not be serviced in realtime, so that probability is now $a+\pi(c)$. The response to this would be to increase $c$ to cover the difference and recall $1/a$ workers for each job instead of 1. However, this is a conservative approach.

A more cost-effective approach would be to alert another worker if the first worker does not respond quickly. If the mean worker response time to an alert is $R$, choose a scalar $\alpha$ and wait until $\alpha R$ for the worker to respond. If the worker has not responded by then, the platform immediately alerts another worker and waits another $\alpha R$ seconds before issuing a third request. There is a constant probability of a worker responding within time $\alpha R$, so the expected number of alerts before getting a response will likewise be a constant. 

Unfortunately, queueing theory cannot easily accommodate this kind of approach. A model including server breakdown is a close match, except that server repair recruits another worker, which means that task arrivals are correlated and no longer Poisson. To bound the expected cost within the queueing theory framework, we envision a more complicated construction, which would be unlikely to be used in a running system. A sketch of the proof follows.  We maintain several tiered retainer pools. If a worker in tier $i$ does not respond within time $\alpha R$, we alert a worker in tier $i+1$. Task arrivals to each queue are now Poisson, since a constant fraction of the requests to tier $i$ will pass through to $i+1$. For example, we might choose $\alpha$ such that half of the requests will respond in time. Then, if tier $i$ has task arrival rate $\lambda$, tier $i+1$ would have arrival rate $\lambda/2$. We would see a sequence of geometrically decreasing pool sizes, meaning the total cost, a geometric sum, will be a small multiple of the first-tier cost, which we have already analyzed.

\section{Improvements to the Retainer Model}
So far, we have analyzed the original form of the retainer model. However, by extending it, we can improve its performance considerably. In this section, we introduce three changes to the retainer model and analyze their impact: retainer subscriptions, globally shared retainer pools with task routing, and predictive recruitment.

\subsection{Retainer Subscriptions}
The worker arrival rate $\mu$ is a limiting factor of the retainer model: previous experiments on MTurk saw $\mu \approx \frac{1}{6}$, or 1 worker every 6 seconds. A small arrival rate means that the retainer pool can take a long time to fill, which is particularly problematic for large bursts or tasks that need multiple simultaneous workers. 

One way to increase $\mu$ is for the platform to put together a panel of \emph{retainer subscribers} who can be directly notified when the retainer pool needs to recruit a replacement. The insight behind this approach is to change from a \emph{pull} model of crowdsourcing, where workers seek out tasks, to a \emph{push} model, where tasks offer themselves to workers. Workers could subscribe to a {\em task type}, so that when the platform needs a retainer worker for a task of that type, the platform could send a dialog notification to one or more subscribed workers and offer them the opportunity to complete one task in the next few minutes. Workers who accept are now on retainer, can continue working on other tasks, and will be interrupted whenever the realtime task arrives.

A push notification is likely to reduce the time it takes to recruit a worker onto retainer, thereby increasing $\mu$.

\subsection{Global Retainer Pools}
In the previous analysis, each requester maintained their own retainer pool. In this section, we analyze how sharing one global retainer pool across requesters improves performance. We also investigate how to route tasks to workers in a globally pooled retainer.

\subsubsection{Global Pool Analysis}
Another way of writing Equation~\ref{eqn:erlang}, the probability of a missed task, is $\pi(c)=\pi(0) \cdot \rho^c/c!$, where $\pi(0) = (\sum_{i=0}^{c}\rho^i/i!)^{-1}$ \cite{gross1998fundamentals}. Recall Stirling's approximation that $c! \approx \sqrt{2\pi c}(c/e)^c$.  Also note that the sum that defines $\pi(0)$ is decreasing geometrically, so we can approximate $\pi(0) \approx e^{-\rho}$, a constant. 
This approximation gives us:
\begin{equation}
\label{eqn:approximation}
\pi(c) \approx e^{-\rho} \sqrt{2\pi c}(e\rho/c)^c
\end{equation}

If we have $k$ different tasks each with traffic intensity $\rho$ and queue size $c$, the probability of an empty pool is roughly $k\pi(c) \approx ke^{-\rho} \sqrt{2\pi c}(e\rho/c)^c$: we multiply by $k$ because each requester independently suffers.

Now suppose we bring all the retainer pools together, creating one ``superpool'' of size $kc$.  The task arrival rate $\lambda$ increases by $k$ but the rate at which we recruit one worker $\mu$ remains unchanged.  Thus the traffic intensity increases by a factor of $k$ to $k\rho$.  So, the probability of an empty pool with combined retainers is
\begin{equation}
\label{eqn:emptycombined}
e^{-k\rho}\sqrt{2\pi kc}(e \rho/c)^{kc} = \sqrt{2\pi kc}\left(e^{-\rho}(e\rho/c)^c\right)^k
\end{equation}
Ignoring the square root factor, we see the main term being \emph{exponentiated} by a factor of $k$.  In other words, the loss rate declines exponentially with the number of retainer pools we bundle.

We can look at some approximations for these results.
Suppose we set $c=(1+\epsilon)\rho$, just above our $c \approx \rho$ knee in the curves from Figure~\ref{fig:equations}.  Then, with a single retainer pool, $\pi(0)$ is about
\begin{align}
e^{-\rho} \sqrt{2\pi c}(e\rho/c)^c &\approx e^{-\rho} (e/(1+\epsilon))^{(1+\epsilon)\rho}\notag\\
&=e^{\epsilon\rho}/(1+\epsilon)^{(1+\epsilon)\rho}\notag\\
&=\left(\frac{e^{\epsilon}}{(1+\epsilon)^{1+\epsilon}}\right)^{\rho}\label{eqn:combinedapprox}
\end{align}

This is the same quantity as shows up in the typical analysis of the upper tail of the Chernoff bound.  There, we generally approximate this quantity as $e^{-\epsilon^2 \rho/3}$, which is reasonably accurate for any $\epsilon < 1$. In short, the probability of an empty pool is roughly $e^{-\epsilon^2 \rho/3}$.

Using this approximation, we can ask what retainer pool size in the \emph{globally shared} case will yield the same empty-pool bound $e^{-\epsilon^2 \rho/3}$ as we found in the singular case.  As we argued above, moving to the globally shared case multiplies
$\rho$ by a factor of $k$.  Since the exponent we care about proportional to $\epsilon^2 \rho/3$, we can decrease $\epsilon$ by a factor of $\sqrt{k}$ and end up with the same bound as the singular case. In other words, the fraction $\epsilon$ of ``buffer'' workers that we need in our retainer pool is proportional to $\sqrt{k}$, as compared to the factor $k$ in the singular case. We thus need many fewer extra workers per extra task: much like standard error decreases by a square root factor as sample size increases, we have less uncertainty in arrival rates as more requesters join together.

\subsubsection{Task Routing}
Shared retainer pools introduce speed and cost improvements, but workers will subscribe to multiple realtime task types and can only work on one realtime task at a time. This situation immediately raises the question of how to decide which worker should be assigned to each retainer pool when a spot opens up. Market forces like task pricing will help solve this problem, but microtask markets like Mechanical Turk are very clustered on a small number of prices (often $2-5$\textcent). Inefficient task routing could lead to logjams where certain tasks cannot find workers. In this section, we demonstrate that a straightforward approach like uniform randomization could lead to extremely slow response times, and we introduce a linear programming solution that optimizes response times across tasks.

Suppose we have a set of task types $T=t_1,\ldots,t_n$, and tasks of type $t_j$ arrive with Poisson distribution and rate $\lambda_j$. Not every worker can complete every task: workers may have only signed up to be on retainer for particular task types, or they may not have the qualifications for all task types. We split workers into groups $w_1,\ldots,w_m$ that are uniquely identified by the tasks that group can complete. So, for example, $w_1$ might represent all the workers who are on retainer for $t_1$, $t_2$, and $t_3$. We say that $W$ is the set of all worker types ($W=w_1,\ldots,w_m$), and that each $w_i$ has a Poisson arrival rate $\mu_i$.

Given a set of task types $T$, a set of worker types $W$, and arrival rates for each, our goal is to assign workers to tasks to maximize the throughput of the system. To do so in steady state, we need to decide how many worker arrivals---more precisely, what portion of the overall arrival rate---from each group should be assigned to each task.   Let us say that the rate at which workers from group $w_i$ should be assigned to tasks of type $t_j$ is $a_{ij}$. These assignments must sum to the total arrival rate of the worker group: $\sum_{j=1}^m a_{ij} = \mu_i$. For example, in our earlier example of $w_1$, if $\mu_1$ = 1, one possible assignment is $a_{11}=.5, a_{12}=.25, a_{13}=.25$.

\begin{figure}
\centering
\resizebox{\columnwidth}{!}{\includegraphics{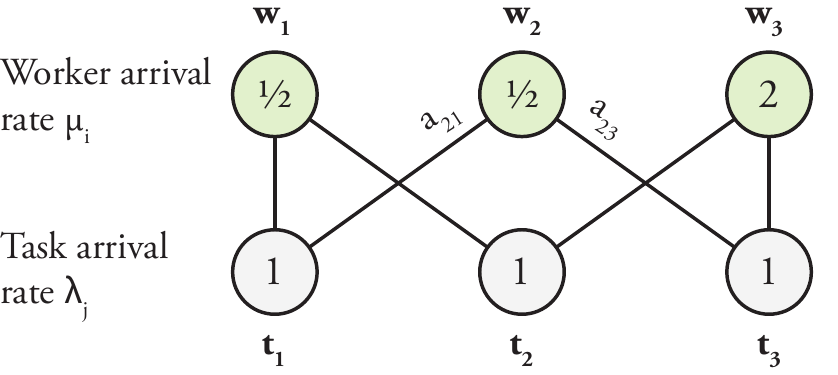}}
\vspace{-20pt}
\caption{A task routing scenario where a typical randomized approach would lead to poor results. $t_1$ would receive relatively few workers. Depending on the values of $\mu_i$, each task type could find itself in this starved state.}
\label{fig:routingbad}
\vspace{-15pt}
\end{figure}

A standard approach would be to assign each worker arrival randomly to one of the task types that he or she can complete. (That is, $a_{ij}$ are equal for any $i$.) However, this approach could result in slow completion times. In Figure~\ref{fig:routingbad}, $w_3$ has four times the arrival rate of $w_1$ or $w_2$. Random assignment would send workers to $t_3$ at rate $1/4 + 1 = 5/4$, whereas $t_1$ would receive workers at just $1/2$. Depending on which workers are online, each of the task types could find itself in a similarly starved state.

Instead, a centralized system can route workers to minimize expected wait time. This goal can be described as a linear programming problem, but in fact can be solved using maximum flow, which is significantly faster than general linear programming. The following constraints suffice to define a linear programming problem --- they indicate that the incoming worker rate to each task type is at least as high as the incoming task rate, and that all the worker assignments from a worker group sum to no more than the arrival rate of that worker group.
\begin{equation}
\begin{aligned}
\sum_i a_{ij} \ge \lambda_j \; & \text{for all } j, \\
\sum_j a_{ij} \le \mu_i \; & \text{for all } i.
\end{aligned}
\end{equation}

We can also choose to be more specific about the quantity to maximize. For example, as we have seen above, task wait times are typically a function of the ratio of arrival rate and service rate ($\lambda / \mu$), known as traffic intensity $\rho$. 
We can define an analogous $\rho$ here to be the ratio of the incoming task arrivals to the summed rate of arrivals from all worker groups for that task: $\rho = \lambda_j / \sum_i a_{ij}$. We then minimize its worst case across all tasks:
\begin{equation}
\begin{aligned}
& \text{minimize } \rho & \\
& \text{subject to} & \rho\sum_i a_{ij} \ge \lambda_j \; & \text{for all } j, \\
& & \sum_j a_{ij} \le \mu_i \; & \text{for all } i.
\end{aligned}
\end{equation}

By merging retainer pools, the platform can thus help guarantee fast results for all tasks.

\subsubsection{Scaling}
One practical difficulty with this approach is estimating $\mu_i$ as the number of task types grows. If there are $|T|$ different task types, there are $2^{|T|}$ different combinations of task types that a worker can sign up for, and thus $|W|=2^{|T|}$. This set is an extremely large number of arrival rates to try and estimate accurately, and will make the linear program hard to solve because there will be an exponential number of constraints. 

However, the problem of efficient feature representation is a common one in machine learning. There are many approaches to this problem.  We may find that in practice only a small number of task type combinations can occur.  We can also enforce this, for example by setting a ceiling on the number of task types a worker can subscribe to at once. With a limit of two subscriptions, $|W|=|T|^2$ instead of $2^{|T|}$.

\subsection{Precruitment: Predictive Recruitment}
Previous research was limited by the length of time it took a worker to respond to the retainer alert. However, our model suggests that even this limit of ``crowds in two seconds'' \cite{twoseconds} is unnecessary, and that crowds could be recruited effectively instantaneously.

The insight behind our solution is \emph{precruitment}: notifying retainer workers before the task actually arrives. The queueing theory model involves estimating $1/\lambda$, the expected length of time before the next task will arrive. If $1/\lambda$ is about the length of time it takes to recall a retainer worker, we can recall a retainer worker and expect to have a task by the time the worker arrives. As we will demonstrate, workers are also happy to wait at a ``Loading...'' screen even if the task is not ready immediately.

Workers take 2-3 seconds to arrive \cite{twoseconds} and will wait for roughly ten seconds afterwards \cite{nielsen1993usability}. The Poisson task arrival process has rate $\lambda$, and Poisson distributions have standard deviation $\sqrt{\lambda}$. So, the platform can precruit $\lambda+\beta \sqrt{\lambda}$ workers per second for upcoming requests, where $\beta$ is a slack variable that controls how many extra standard deviations to precruit for safety. Any workers who do not have tasks within a predetermined wait time would need to be paid and dismissed. However, as the platform becomes large, the standard deviation will become proportionally smaller relative to the mean, making it possible to waste very little money on extra workers.

In fact, the entire precruitment system can be represented as its own M/M/c/c queueing system. Many of the same techniques introduced earlier can be applied to help optimize the size of a precruitment pool in relation to the standard retainer pool.

\subsection{Evaluation}
We ran a study on Mechanical Turk as a proof-of-concept for precruitment. In the study, we followed the protocol of \citeasnoun{twoseconds} by offering three cents for a one-minute retainer task: a game of Whack-a-Mole. After waiting on retainer for one minute, workers responded to the retainer alert and were asked to quickly click on the picture of a mole randomly placed in a 3x3 grid of dirt mounds. However, after responding to the alert and before the mole appeared, workers needed to wait for a randomly selected length of time between 0 and 20 seconds while a ``Loading...'' indicator displayed.

We measured the length of time between the appearance of the mole and: a) mouse movement in the direction of the mole, and b) the click on the mole. We discarded any responses where worker clicked on a dirt mound instead of the mole or where the browser did not record millisecond-precision timing. After filtering, our dataset consisted of fifty workers who completed N=373 Whack-a-Mole tasks. One limitation of our design is that Whack-a-Mole is a relatively enjoyable task, and workers might not be so attentive for less game-like tasks. 

\begin{figure}
\centering
\vspace{-5pt}
\resizebox{\columnwidth}{!}{\includegraphics{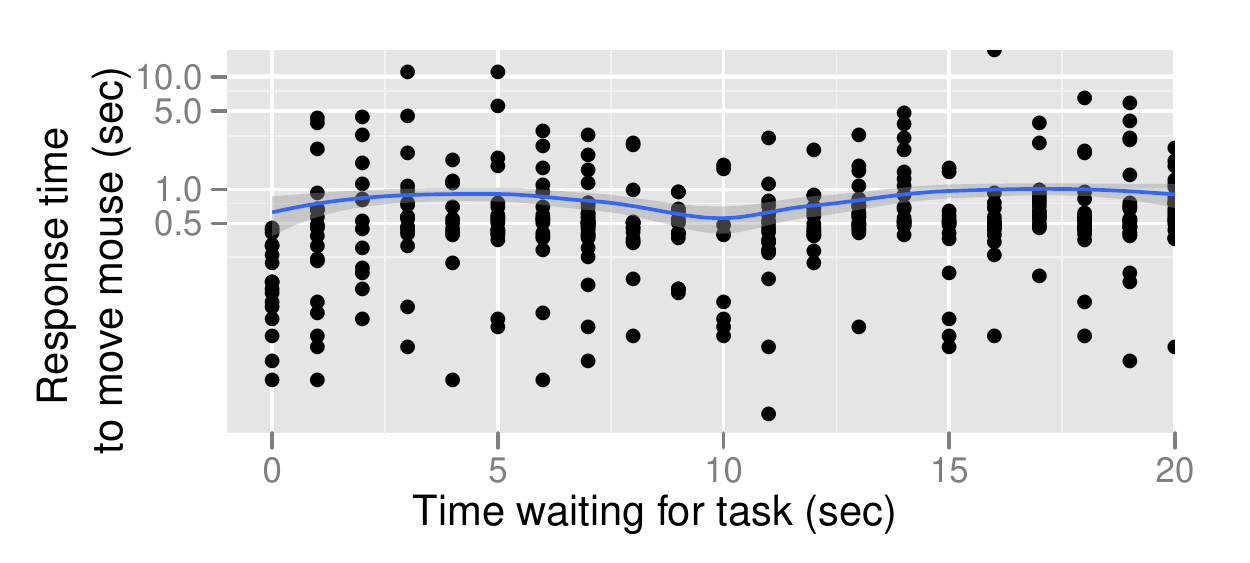}}
\vspace{-25pt}
\caption{The median length of time between the mole image appearing and the workers moving to click on it was 0.50 seconds. So, a platform can recall retainer workers early and get crowds in half a second instead of waiting for the workers to respond to the retainer alert.}
\label{fig:precruitment}
\vspace{-15pt}
\end{figure}

The median length of time between the mole's appearance and the worker moving the mouse toward the mole to click on it was 0.50 seconds across all wait times (mean 0.86, std. dev. 1.45, Figure~\ref{fig:precruitment}). The median length of time before clicking on the mole was 1.12 seconds (mean 1.87, std. dev. 2.23). There is a negligible correlation between wait time and mouse movement delay ($R^2=.001$), suggesting that workers react roughly as quickly right after they arrive as they do twenty seconds later. 

We can use the same dataset to compare precruitment to the retainer approach presented in \citeasnoun{twoseconds}. This comparison is possible because a ``Loading...'' delay of zero seconds is the exact same worker experience as the standard retainer model. We are interested in the lag time between the task arriving and mouse movement to whack the mole. Here, a new task results in an alert being sent to the worker, so we start our timer with the alert. Without precruitment, the median time between task posting and mouse move was 1.36 seconds (mean 1.41, std. dev. 0.30). 

This result suggests that, had we used a standard retainer model with this task, we would have seen mouse movement typically after 1.36 seconds. Using precruitment, we get mouse movement in 0.5 seconds. Precruitment finally breaks through the sub-second cognitive barrier that keeps users in flow \cite{nielsen1993usability}.

\section{Discussion}
Our model has several limitations. One limitation is that an M/M/c model may be a better match for certain retainer implementations where the idea is to handle tasks FIFO and not immediately give up on realtime response for tasks when the pool is empty. Second, worker recall delays depend on the length of time the worker has been waiting on retainer \cite{twoseconds}, but our analysis ignores this fact. Third, our model assumes that it can always recruit new retainer workers into the pool, but the retainer population is limited in practice. However, we believe that these observations can be integrated into our optimizations.

One empirical question we have not addressed is the number of workers that need to be on a crowdsourcing platform to make sure that requesters can maintain full retainer pools. This number also depends on the percentage of workers who are willing to sign up for retainer tasks. Since the retainer model pays more than batch tasks, we anticipate that this percentage will be high. On Mechanical Turk, our experience is that it is not difficult for a single requester to get twenty or thirty workers on retainer simultaneously. However, as more requesters use retainers, these dynamics may shift.

While realtime retainers are the motivating example in this paper, the entire Mechanical Turk platform can be thought of as a large retainer system where workers are paid zero retainer wage and the worker recall rate is extremely slow, since workers return on their own initiative rather than by recall. Precruitment is another kind of retainer model queue where workers are recalled before the task even arrives. All three queues could be analyzed together as a queueing network in order to more effectively understand the entire system. However, it is also possible to bound the probability of a slow task response via the probability that any of the retainer pools are empty. Supported by our results so far, we suggest that queueing theory can be applied for many other problems in the space of realtime crowdsourcing as well.

Our analysis suggests that paid crowdsourcing platforms could integrate a globally-managed retainer into their design. This will not only change the types of crowdsourcing that are common, but will also introduce new elements of worker reputation. We suggest two new reputation statistics. First, a worker's median response time characterizes how quickly they respond to the alert and begin working on a retainer task. Requesters prefer workers with low response times. Second, workers are tagged with a response rate: the percentage of the time that they successfully respond to a retainer alert. If a worker does not respond to the alert within a given length of time (e.g., five seconds), the system finds another person and the worker is not paid. 

\section{Conclusion}
In this paper, we have analyzed and optimized the retainer model for realtime crowdsourcing. We applied queueing theory to demonstrate specific relationships between task and worker arrival rates, the size of the retainer pool (workers waiting for a task), cost and expected wait time. We introduced a technique for choosing the optimal retainer pool size given a requester's needs, and for integrating abandonment into the queueing theory model. Finally, we described three new techniques that improve the performance of the retainer model: retainer subscriptions, shared retainer pools, and \emph{precruitment}, or recalling retainer workers before the task arrives. These techniques suggest directions for future platform development, and have already shown promise in returning results to users in 500 milliseconds.

\bibliography{realtime_model}
%

\end{document}